\documentstyle[sprocl]{article}
\input{psfig}
\def\Journal#1#2#3#4{{#1} {\bf #2}, #3 (#4)}

\def\NPB{{\em Nucl. Phys.} B}
\def\PLB{{\em Phys. Lett.}  B}
\def\PRL{\em Phys. Rev. Lett.}
\def\PRD{{\em Phys. Rev.} D}
\def\ZPC{{\em Z. Phys.} C}
\def\ZPR{\em ZhETF Pis. Red.}
\def\JETPL{\em JETP Lett.}
\def\PTP{\em Prog. Theo. Phys.}
\def\ARNPS{\em Ann. Rev. Nucl. Part. Sci.}
\def\IJMPA{{\em Int. J. Mod. Phys.} A}

\def\ra{\rightarrow}

\def\be{\begin{equation}}
\def\ee{\end{equation}}
\def\bea{\begin{eqnarray}}
\def\eea{\end{eqnarray}}

\def\O{{\cal O}}

\def\epsK{\varepsilon_K}

\def\Re{{\rm Re}}
\def\Im{{\rm Im}}
\def\ra{\rightarrow}
\def\gsim{{~\raise.15em\hbox{$>$}\kern-.85em
          \lower.35em\hbox{$\sim$}~}}
\def\lsim{{~\raise.15em\hbox{$<$}\kern-.85em
          \lower.35em\hbox{$\sim$}~}}

\def\aPK{a_{\psi K_S}}

\def\aFK{a_{\phi K_S}}

\begin{document}

\begin{titlepage}
 
\begin{flushright}
WIS-97/28/Sep-PH\\
hep-ph/9709301
\end{flushright}
 
\vspace{1.5cm}
 
\begin{center}
\Large\bf Recent Developments in\\
Theory of CP Violation
\end{center}
 
\vspace{1.2cm}
 
\begin{center}
Yosef Nir\\
{\sl Department of Particle Physics,
 Weizmann Institute of Science,\\
 Rehovot 76100, Israel}
\end{center}
 
\vspace{1.3cm}

\begin{center}
{\bf Abstract}\\[0.3cm]
\parbox{11cm}{
Four topics in theory of CP violation are reviewed. (a) CP violation
in $B$ decays: We describe a new clean way of constraining the angle
$\gamma$ of the unitarity triangle and how new CP violation in decay
amplitudes can signal new physics. (b) CP violation in $K$ decays:
We explain the special features of the decay $K_L\ra\pi^0\nu\bar\nu$
both as a measurement of Standard Model CP violating parameters
and as a probe of new physics. (c) CP violation in $D$ decays:
We describe the consequences of CP violation from new physics
in $D-\bar D$ mixing. (d) CP violation in Supersymmetry: We
explain how a combination of measurements of CP violating processes
will give insight into the flavor and CP structure of supersymmetry.}
\end{center}
 
\vspace{1cm}
 
\begin{center}
{\sl To appear in the Proceedings of the\\
18th International Symposium on Lepton Photon Interactions\\
Hamburg, Germany, July 28 -- August 1 1997}
\end{center}
 
\vspace{1.5cm}
 
\vfil
\noindent
WIS-97/28/Sep-PH\\
September 1997
 
\end{titlepage}

\newpage
 
\setcounter{page}{1}

\title{RECENT DEVELOPMENTS IN THEORY OF CP VIOLATION}
\author{ YOSEF NIR }
\address{Department of Particle Physics,
 Weizmann Institute of Science,\\ Rehovot 76100, Israel}
\maketitle\abstracts{
Four topics in theory of CP violation are reviewed. (a) CP violation
in $B$ decays: We describe a new clean way of constraining the angle
$\gamma$ of the unitarity triangle and how new CP violation in decay
amplitudes can signal new physics. (b) CP violation in $K$ decays:
We explain the special features of the decay $K_L\ra\pi^0\nu\bar\nu$
both as a measurement of Standard Model CP violating parameters
and as a probe of new physics. (c) CP violation in $D$ decays:
We describe the consequences of CP violation from new physics
in $D-\bar D$ mixing. (d) CP violation in Supersymmetry: We
explain how a combination of measurements of CP violating processes
will give insight into the flavor and CP structure of supersymmetry.}
 
\section{Introduction}
 
It is often said that the subject of CP symmetry and its violation
is one of the least understood in particle physics.
A better statement would be to say that it is experimentally one of
the least constrained. CP violation is an expected consequence of
the Standard Model with three quark generations, but is one of the least
tested aspects of this model. The only part of CP violation that,
at present, is considered puzzling by theorists is the lack of
CP violation in strong interactions, that is the strong CP problem.
The CP violation that shows up in a small fraction of weak decays is
accommodated simply in the three-generation Standard Model Lagrangian.
All it requires is that we do not impose CP as a symmetry.
 
However, while we know that CP violation occurs, because it has been
observed in $K$ decays,\cite{CCFT}
we do not yet know whether the pattern of CP violation predicted by the
minimal Standard Model is the one found in nature. The $K$-decay
observations, together with other measurements, place constraints on the
parameters of the Standard Model mixing matrix (the CKM matrix
\cite{Cabi,KoMa}) but do not yet provide any test.  A multitude of
large CP-violating effects are expected in various $B$ decays and in
$K\ra\pi\nu\bar\nu$ decays, some of which are very cleanly predicted by
the Standard Model.
If we can make enough such independent observations
then it will be possible to test the Standard Model predictions for
CP violation. Either we will see that the relationships between various
measurements are consistent with the Standard Model predictions and fully
determine the CKM parameters or we will find that there is no single
choice of CKM parameters that is consistent with all measurements.
 
This latter case, of course, would be much more interesting.
It would indicate that there is a contribution of physics beyond the
Standard Model. There may be enough information in the pattern of the
inconsistencies to tell us something about the nature of the new physics
contributions. Thus the aim of the game is to measure enough quantities
to impose redundant constraints on Standard Model parameters, including
particularly the convention independent combinations  of CP-violating
phases of CKM matrix elements.
 
There are also many CP violating observables where the Standard Model
contributions are too tiny to be observed. Most noticeable
among these are the electric dipole moments of the neutron and the
electron, CP violation in top production and decay, CP violation
in $D-\bar D$ mixing, and transverse lepton polarization in meson decays.
If experiments find a signal then, again, this will indicate new physics.
The pattern of CP violation is likely to provide useful information on
the details of the relevant new physics.
 
One may well ask, after the many successes of the Standard Model, why
we would expect violations to show up in such a low-energy regime. The
best answer is simply that it has not yet been tested. Theorists
will give a variety of further reasons. Many  extensions of the
Standard Model have additional sources of CP violating effects, or
effects which change the relationship of the measurable quantities to the
CP-violating parameters of the Standard Model.
 
In addition there is one great puzzle in cosmology that relates to CP
violation, and that is the disappearance of the antimatter.\cite{Sakh}
In grand unified theories, or even in the Standard Model at sufficiently
high temperatures, there are baryon number violating processes. If such
processes are active then thermal equilibrium produces equal
populations of particles and antiparticles. Thus in modern theories of
cosmology the net baryon number of the universe is zero in the early
high temperature epochs. Today it is clearly not zero, at least in our
local region. We will not here give a full discussion of the cosmological
arguments. It suffices to remark that there is a large class of theories
in which the baryon number asymmetry is generated at the weak phase
transition.\cite{CKN}
Such theories, however, must include CP violation from sources beyond
the minimal Standard Model. Calculations made in that model show that it
does not generate a large enough matter-antimatter imbalance to produce
the baryon number to entropy ratio observed in the universe today. This
is a hint that CP violation from beyond Standard Model sources is worth
looking for. It is by no means a rigorous argument. There are theories in
which baryon number is generated at a much higher temperature and then
protected from thermalization to zero by $B-L$ (baryon number minus
lepton number) symmetry. Such theories do not in general require any new
low energy CP violation mechanism. Neither do they forbid it.
 
More generally, since we know there is CP violation in part of the
theory, any extension of the Standard Model cannot be required to be CP
symmetric. Any additional fields in the theory bring possible additional
CP violating couplings. Even assumptions such as soft or spontaneous
CP symmetry breaking leave a wide range of possibilities.
Further experimental constraints, from experiments such as the
$B$ factory, are needed.
 
In this talk, we will focus on four aspects of CP violation:
 
(i) {\it CP violation in $B$ decays}.\cite{BuFla}$^-$\cite{SiSi}
Within the Standard Model framework, we describe a
new method to constrain the angle $\gamma$ of the unitarity triangle that
is theoretically clean and experimentally feasible. Beyond the
Standard Model, we explain how CP violation in the decay amplitudes can
be useful for discovering new physics.
 
(ii) {\it CP violation in $D$ decays}.\cite{BNS}$^-$\cite{CKLN}
We study the neutral $D$
decays into final $K^\pm\pi^\mp$. We explain how CP violation from
New Physics can affect the search for mixing through this decay.

(iii) {\it CP violation in $K$ decays}.\cite{MaPa}$^-$\cite{BuBuEW}
We focus on the $K_L\ra\pi^0\nu\bar\nu$ decay.
Within the Standard Model, it gives a clean measurement of the
CP violating parameter $\eta$. Beyond the Standard Model, it probes new
CP violating phases in the $s\ra d\nu\bar\nu$ decay.
  
(iv) {\it CP violation as a probe of Supersymmetry}.
\cite{PoTo}$^-$\cite{KoLi}\ \cite{DDO,BaSt,CKLN} We describe
the various developments in understanding the flavor and CP
problems in Supersymmetry. We explain how measurements of CP violation
could distinguish among the various solutions to these problems.
 
Unfortunately, due to lack of time and space, we have to leave out
many other topics where there have been recent interesting
developments, {\it e.g.} CP violation in top and Higgs physics,
\cite{ABB}$^-$\cite{BAS} CP violation in neutrino oscillations,
\cite{Tani}$^-$\cite{KoSa} baryogenesis,\cite{CRV}$^-$\cite{MuVi}
and various others.\cite{Nels}$^-$\cite{BLP}
 
\section{CP Violation in Meson Decays}
 
\subsection{Formalism}
 
To establish our notations and to understand similarities
and differences between $K$, $D$ and $B$ decays, we here briefly
review the formalism of CP violation in meson decays.
 
We define decay amplitudes $A_f$ and $\bar A_f$ through
\be
A_f=\langle f|H|B^0\rangle,\ \ \ \bar A_f=\langle f|H|\bar B^0\rangle.
\label{defAf}
\ee
We denote by $p$ and $q$ the components of the interaction eigenstates
in the neutral meson mass eigenstates:
\be
|B_{1,2}\rangle\ =\ p|B^0\rangle\pm q|\bar B^0\rangle.
\label{defpq}\ee
Finally, the complex quantity $\lambda_f$ is defined by
\be
\lambda_f = {q\over p}{\bar A_f\over A_f}.
\label{deflam}\ee
 
The possible manifestations of $CP$ violation can be classified in a
model independent way:
\begin{description}
\item[$(i)$] {CP violation in decay, which occurs in both charged and
neutral decays, when the amplitude for a decay and its $CP$-conjugate
process have different magnitudes:
\be
|\bar A_{\bar f}/A_f|\neq1.
\label{CPdec}\ee
($\bar f$ denotes the CP-conjugate of the state $f$.)}
\item[$(ii)$] {CP violation in mixing, which occurs when the two neutral
mass eigenstate admixtures cannot be chosen to be $CP$-eigenstates:
\be
|q/p|\neq1.
\label{CPmix}\ee}
\item[$(iii)$] {CP violation in the interference between decays with and
without mixing, which occurs in decays into final states that are common
to $B^0$ and $\bar B^0$.  It often occurs in combination with the other
two types but there are cases when, to an excellent
approximation, it is the only effect, namely
\be
\Im\lambda_f\neq0\ \ \ (|\lambda_f|\approx1).
\label{CPint}\ee}
\end{description}
 
\subsection{The CKM Constraints}
 
To understand the Standard Model predictions for CP asymmetries
in various neutral meson decays, we study the constraints on
the CKM parameters from $|V_{cb}|$, $|V_{ub}/V_{cb}|$,
$\Delta m_{B_d}$, $\epsK$ and $\Delta m_{B_s}$.
We use a new method of statistically combining the many measurements
involving CKM parameters.\cite{PlSc} This method
was adopted by the BaBar collaboration \cite{BaBar}
and is described in detail in \cite{GNPS}.
 
There are two types of errors which enter the determination of the CKM
parameters: experimental errors and uncertainties due to theoretical
model dependence. These two types of errors will be treated differently.
Experimental errors are generally assumed to be
Gaussianly distributed and can then enter a $\chi^2$ test.
For the quantities with Gaussian errors, we use \cite{Schn,PDG}
\bea
|V_{cb}|&=&0.039\pm0.004,\nonumber \\
|V_{ub}/V_{cb}|_{\rm exp}&=&|V_{ub}/V_{cb}|_T\pm0.05,\nonumber \\
\Delta m_{B_d}&=&0.463\pm0.018\ {\rm ps}^{-1},\nonumber \\
|\epsK|&=&(2.258\pm0.018)\times10^{-3}.
\label{Eerrors}\eea
(The subscript $T$ implies that we here
refer to the hadronic model dependent range for $|V_{ub}/V_{cb}|$ to
which an experimental error should be added to give the full
uncertainty.)
A large part of the uncertainty in translating the experimental
observables to the CKM parameters comes, however, from errors related to
the use of hadronic models. At present, one cannot
assume any shape for the probability density of these quantities
(certainly not Gaussian) and include it in the fit. We thus do not
assume any shape for these distributions but use a whole set of
`reasonable' values for the parameters. Specifically, we scan the ranges
\bea
0.06\leq&|V_{ub}/V_{cb}|_T&\leq0.10,\nonumber \\
160\leq &f_{B_d}\sqrt{B_{B_d}}&\leq240\ MeV,\nonumber \\
0.6\leq &B_K&\leq1.0.
\label{Terrors}\eea
The mass difference in the $B_s$ system has not been measured
and only 95\% CL limits have been obtained:\cite{Schn}
\be
\Delta m_{B_s}\ge10.0\ {\rm ps}^{-1}.
\label{Bsbound}\ee
Such a limit is only a small part of the information and it cannot
be included directly in the $\chi^2$ minimization.
In our analysis, we include the full information from
the amplitude method that is now being used by the LEP
$\Delta m_{B_s}$ averaging Working Group.\cite{Schn}
We also use \cite{Sach}
\be
{B_{B_s}f_{B_s}^2\over B_{B_d}f_{B_d}^2}=1.30\pm0.18.
\label{xirange}\ee
 
\begin{figure}
\centerline{$(a)$}
\centerline{
\psfig{file=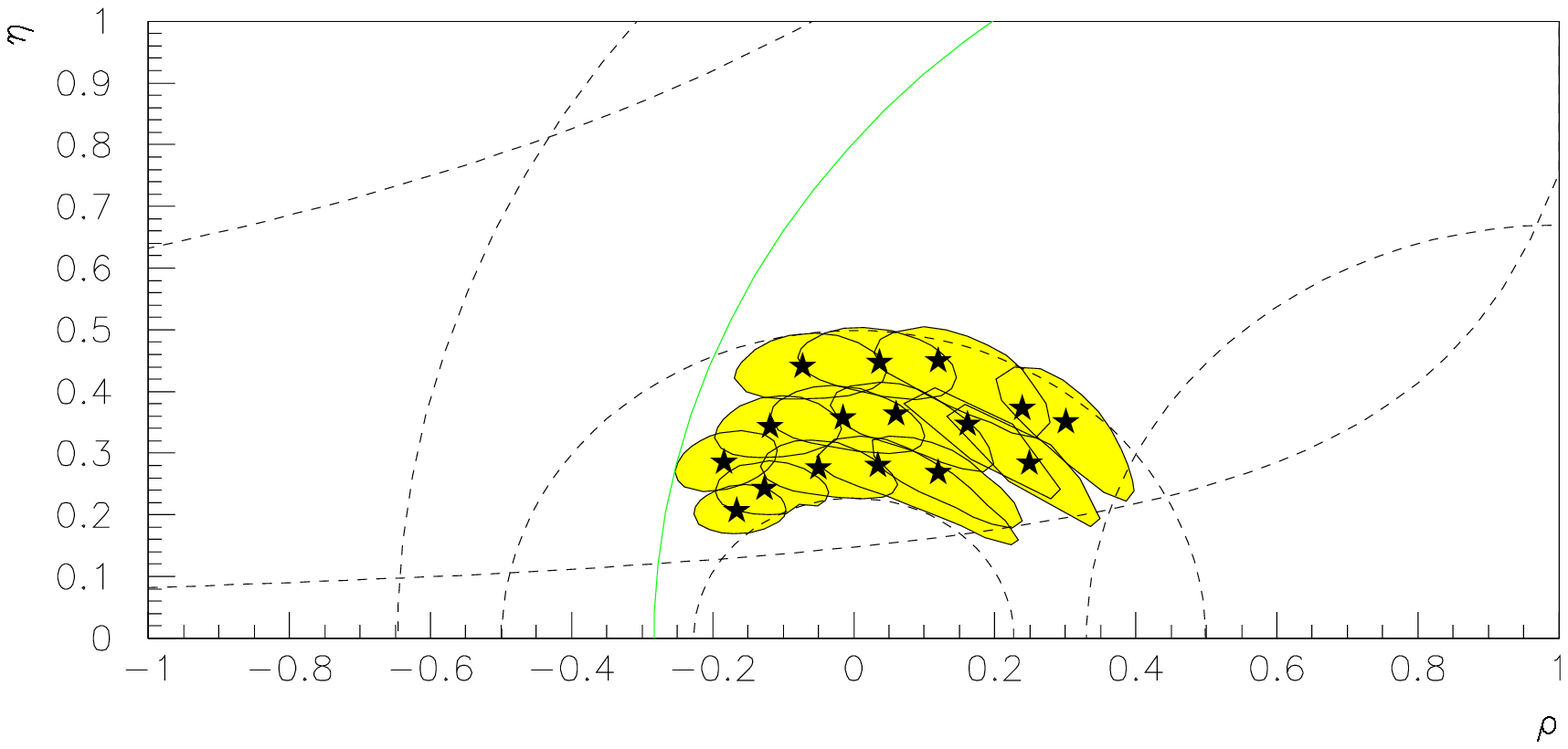,width=370pt,bbllx=0pt,bblly=410pt,bburx=612pt,bbury=660pt
}}
\centerline{$(b)$}
\centerline{
\psfig{file=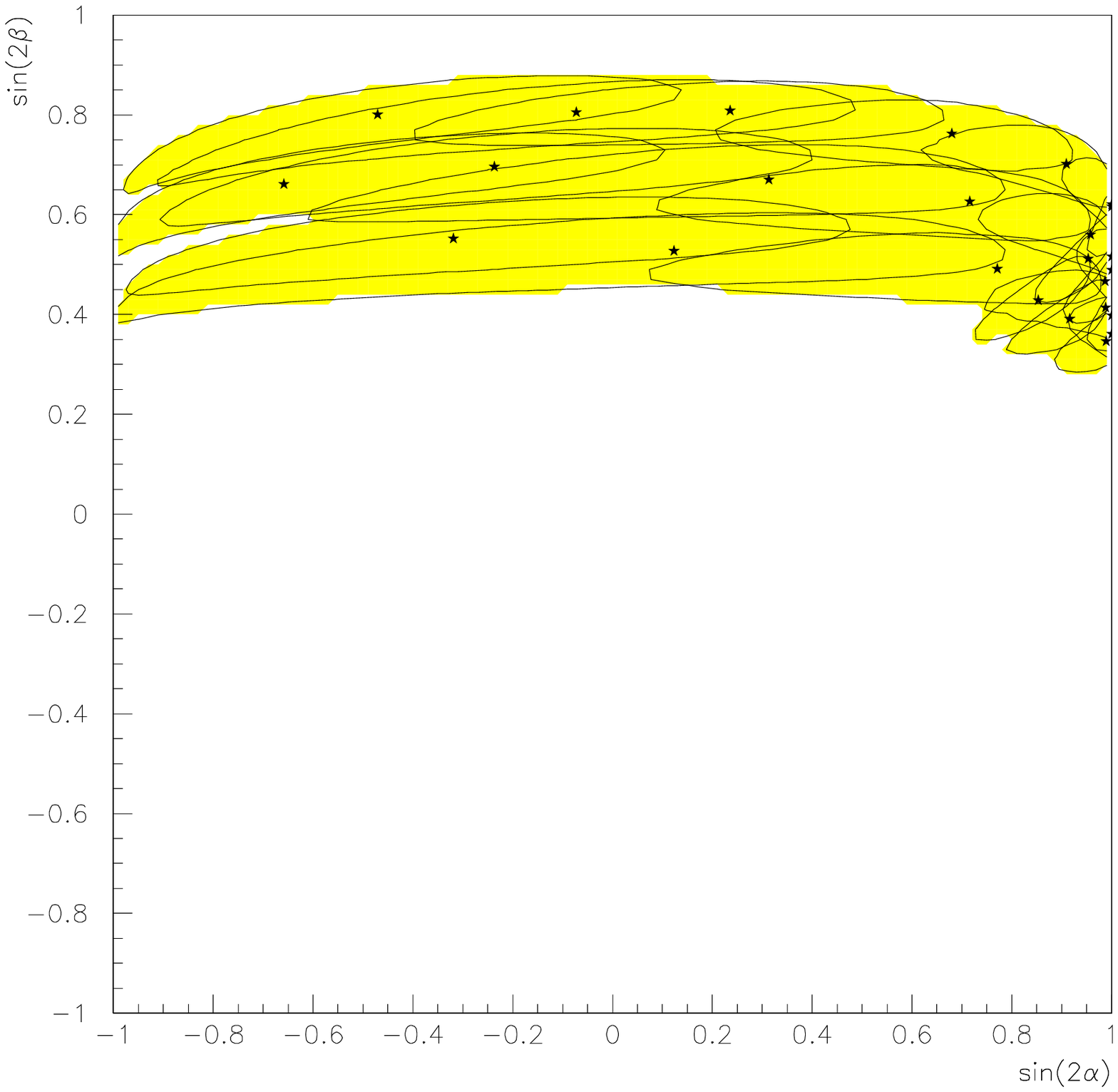,width=320pt,bbllx=0pt,bblly=160pt,bburx=612pt,bbury=653pt
}}
\caption{The present allowed range $(a)$ in the $\rho-\eta$ plane 
and $(b)$ in the $\sin2\alpha-\sin2\beta$ plane
using constraints from $|V_{cb}|$,
$|V_{ub}/V_{cb}|$, $\Delta m_{B_d}$, $\epsK$ and $\Delta m_{B_s}$.
For the methods used in this analysis, see refs. (160,161,162).
\label{presentRE}}
\end{figure}

The present allowed region at 95\% CL in the
$\rho-\eta$ plane is presented in Fig. \ref{presentRE}(a).
Another useful presentation is in the $\sin2\alpha-\sin2\beta$ 
plane.\cite{SoWo,NiSa} The present allowed region at 95\% CL 
is shown in Fig. \ref{presentRE}(b).
 
Examining the figures, we find that, if the theoretical parameters
are within the range (\ref{Terrors}), the following ranges for
the various angles of the unitarity triangle are allowed at the
95\% CL:
\bea
0.28\le&\sin2\beta&\le0.88,\nonumber\\
-1.0\le&\sin2\alpha&\le1.0,\nonumber\\
0.23\le&\sin^2\gamma&\le1.0.
\label{SMphases}\eea 
 
\section{$B$ Physics}
 
A huge amount of work has been devoted to CP violation in $B$ decays.
This is no doubt a result of the forthcoming $B$-factories, BaBar
and Belle. The effort goes in two main directions: how to determine
best the values of the CP violating angles of the unitarity triangle
and how to find New Physics. Instead of trying to review all the
work that has been done in this field, I will give two examples
of recent attractive developments. In the direction of measuring
CKM phases, I will describe a new method to constrain $\gamma$.
In the direction of exploring new physics, I will describe a
method that uses possible new phases in the decay amplitudes
(rather than in the mixing).
 
\subsection{Constraining $\gamma$}
 
Of the three angles of the unitarity triangle, $\gamma$ is the
most difficult one to measure in a $B$-factory. Many clever methods
were suggested, but most of them either suffer from rather large
hadronic uncertainties or are very difficult, not to say impossible,
to carry out in a $B$-factory. Two methods, however, are
theoretically rather clean. One is a proposal by Atwood, Dunietz
and Soni,\cite{ADS} based on an idea by Gronau and Wyler,\cite{GrWy}
using triangle relations in $B\ra D^0K$ decays. The other,
which is described in detail below, was proposed by
Fleischer and Mannel:\cite{FlMag} using the branching
ratios of four $B\to\pi K$ decay modes, it is possible to derive a bound
on the angle $\gamma$ of the unitarity triangle which,
under certain circumstances, is free of hadronic uncertainties.
 
The amplitudes for the relevant $B\to\pi K$ decays can be
written as follows:
\bea
A(B^0\ra\pi^- K^+) &=& A_c^0 - A_u^0 e^{i\gamma} e^{i\delta},\nonumber\\
A(\bar B^0\ra\pi^+K^-) &=& A_c^0 - A_u^0 e^{-i\gamma} e^{i\delta},
\nonumber\\
A(B^+ \ra \pi^+ K^0) &=& A_c^+ - A_u^+ e^{i\gamma} e^{i\delta'},
\nonumber\\
A(B^- \ra \pi^- \bar K^0) &=& A_c^+ - A_u^+ e^{-i\gamma} e^{i\delta'}.
\label{fourmodes}\eea
The following two assumptions are very likely to hold with regard
to these four channels:
 
{\it 1. The contributions to $A_u$ that do not come from tree
amplitudes can be neglected}.\cite{FleiRev} The reason is that the
penguin amplitudes contributions to $A_u$ are suppressed compared
to their contributions to $A_c$ by $\O(|V_{ub}V_{us}|/|V_{tb}V_{ts}|)
\sim0.02$. Then in the charged $B$ decays, which require a
$b\to d\bar ds$ transition, we can neglect $A_u$ while in the neutral $B$
decays, which can also be mediated by a $b\to u\bar u s$ transition, we
take into account only the tree amplitude $A_T$:
\be
A_u^+=0,\ \ \ A_u^0=A_T.
\label{AuAT}\ee
 
{\it 2. The contributions from electroweak penguins can be 
neglected.}\cite{FleiRev} Indeed these contributions can be 
reliably estimated and
they are expected to be $\O(0.01)$ of the leading contributions.
Then $A_c$ comes purely from QCD penguin amplitudes $A_P$
which, as a result of the $SU(2)$ isospin symmetry of the strong
interactions, contribute equally to the charged and neutral $B$ decays:
\be
A_c^0 = A_c^+ = A_P.
\label{AcAP}\ee
We define
\bea
r&\equiv& A_T/A_P,\nonumber\\
\Gamma(B_d\ra\pi^\mp K^\pm)&\equiv&{\Gamma (B^0\ra\pi^-K^+)+\Gamma
(\bar B^0\ra\pi^+ K^-)\over 2},\nonumber\\
\Gamma(B^\pm\ra\pi^\pm K)&\equiv&{\Gamma(B^+\ra\pi^+K^0)+\Gamma
(B^-\ra\pi^-\bar K^0)\over2},\nonumber\\
R&\equiv&{\Gamma(B_d\ra\pi^\mp K^\pm)\over\Gamma(B^\pm\ra\pi^\pm K)}.
\label{ratios}\eea
With the two approximations (\ref{AuAT}) and (\ref{AcAP})
one gets \cite{FlMag}
\be
R=1-2r\cos\gamma\cos\delta + r^2.
\label{defR}\ee
In general, constraints on $\gamma$ from eq. (\ref{defR})
depend on hadronic
physics. In particular, while $R$ is a measurable quantity, $r$
and $\cos\delta$ are hadronic, presently unknown parameters.
(We treat $r$ as a free parameter. Estimates based on factorization
and on $SU(3)$ relations prefer $r\lsim 0.5$.\cite{FlMag})
Fortunately, one can find an inequality that is independent of $r$
and $\cos\delta$:\cite{FlMag}
\be
\sin^2\gamma \leq R.
\label{Limit}\ee
 
\begin{figure}
\centerline{$(a)$}
\centerline{
\psfig{file=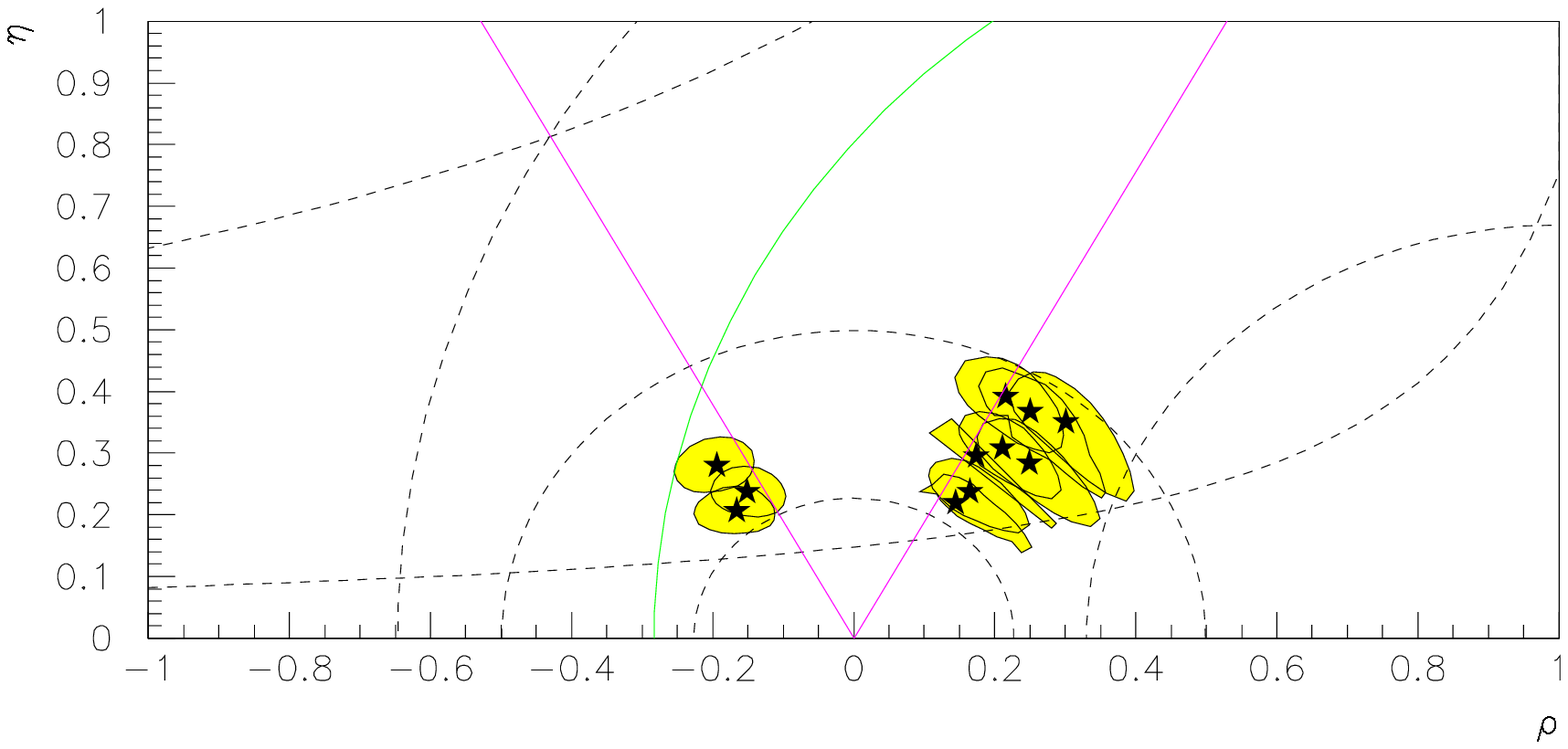,width=370pt,bbllx=0pt,bblly=410pt,bburx=612pt,bbury=660pt
}}
\centerline{$(b)$}
\centerline{
\psfig{file=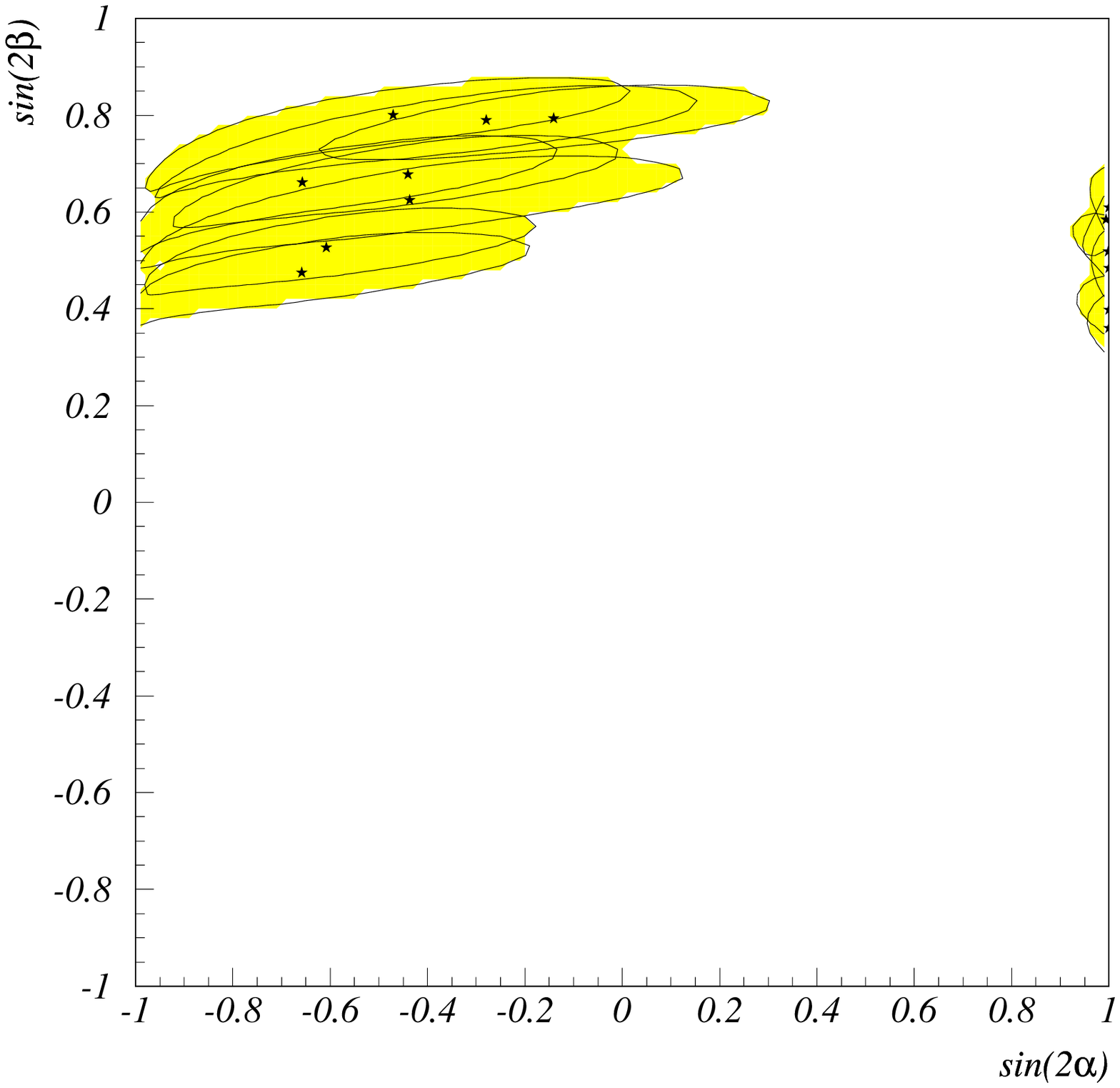,width=320pt,bbllx=0pt,bblly=160pt,bburx=612pt,bbury=653pt
}}
\caption{The effect of the FM bound with $R=0.65\pm0.08$
on the constraints $(a)$ in the $\rho-\eta$ plane and
$(b)$ in the $\sin2\alpha-\sin2\beta$ plane. The central value
for $R$ is taken from the present CLEO measurement (ref. (169))
while the error is our estimate for the accuracy that can be
obtained with about 80 fb$^{-1}$ in $B$-factories.
For all other constraints, we use present data.
\label{FMRE}}
\end{figure}

Clearly, the bound (\ref{Limit}) is significant only for $R<1$.
Recent CLEO results\cite{CLEOR}
give $R=0.65\pm0.40$. Thus, we may be fortunate and indeed have $R<1$.
As soon as an upper bound on $R$ below unity is obtained, the limit
(\ref{Limit}) will give, within the Standard Model, useful constraints in
the $\rho-\eta$ (fig. \ref{FMRE}(a)) and $\sin2\alpha-\sin2\beta$
(fig. \ref{FMRE}(b)) planes. It can also probe new physics.\cite{FlMaNP}
 
\subsection{New CP Violation in Decay Amplitudes}
 
Grossman and Worah\cite{GrWo} have argued that new CP violating effects
in $\Delta B=1$ processes can be cleanly signalled in experiment
even if the effects are smaller than the widely discussed new
CP violation in $\Delta B=2$ processes. The reason is that to see
the decay effects, one compares two experimentally measured quantities,
and does not need to know the theoretically allowed range for
either of them.
 
To explain the main points, we take the explicit example of the
CP asymmetries in $B\ra\psi K_S$ and $B\ra\phi K_S$, which we denote
by $\aPK$ and $\aFK$, respectively. Within the Standard Model,
each of these is dominated by a single CKM phase. Consequently,
to a very good approximation, the source of the CP asymetries
is CP violation in the interference of decays with and without
mixing, namely $\Im\lambda\neq0$. Furthermore, the asymmetries
can be calculated in a theoretically clean way, giving
\be
\aPK=\sin2\beta,\ \ \ \aFK=\sin2\beta,
\label{SMaPK}\ee
so that the present accuracy of the Standard Model prediction
for these asymmetries is given by (see fig. 2):
\be
0.3\lsim\sin2\beta\lsim0.9\ \ (95\%\ {\rm CL}).
\label{stwob}\ee
The Standard Model relation (\ref{SMaPK}) for $\aPK$ is extremely clean.
For $\aFK$, effects of $O(|V_{ub}V_{us}|/|V_{tb}V_{ts}|)\lsim0.03$
are neglected. We thus learn that the Standard Model predicts \cite{GrWo}
\be
\aPK=\aFK\ {\rm to\ within}\ 6\%.
\label{SMFP}\ee
 
Most studies of new physics effects on CP asymmetries in neutral
$B$ decays have focussed on new CP violation in $B-\bar B$
mixing. (For recent, model independent studies of this case, see
\cite{DDO,SiWo,GNW}.) The strong suppression of the Standard
model box diagrams by the fourth order of the weak coupling and
small CKM angles indeed allows for competing, maybe even dominant
contributions from new physics. In this case, one can parameterize
the new physics effects by two new parameters, $r_d$ and $\theta_d$,
defined by
\be
{\langle B^0|{\cal H}_{\rm eff}^{\rm full}|\bar B^0\rangle\over
\langle B^0|{\cal H}_{\rm eff}^{\rm SM}|\bar B^0\rangle}=
(r_d e^{i\theta_d})^2.
\label{defrtd}\ee
The important features in this framework are that large effects
on CP asymmetries in $B^0$ decays are possible and that the asymmetries
are shifted universally. The shift depends on the new CP violating
phase $\theta_d$ only. In particular:
\be
\aPK=\sin2(\beta+\theta_d),\ \ \ \aFK=\sin2(\beta+\theta_d),
\label{NPaPK}\ee
and the equality between the asymmetries (\ref{SMFP}) is maintained.
The angle $\theta_d$ is generally unconstrained. If indeed $\sin2\beta
\sim0.6$, then a rather large $\theta_d$ is required in order that
the deviation from the Standard Model range (\ref{stwob}) will be
manifest.
 
As for the decay amplitudes, the $B\ra\psi K_S$ decay goes through
the quark $\bar b\ra\bar s c\bar c$ transition which gets
contributions from Standard Model tree diagrams with only mild
CKM suppression. It is then very unlikely that new physics could
affect this decay in a significant way. On the other hand, the
$B\ra\phi K_S$ decay goes through the quark $\bar b\ra\bar s s\bar s$
transition. This is a FCNC process to which the leading Standard Model
contributions are QCD penguin amplitudes with an extra suppression by
$\alpha_s$ and a loop factor. Here one could easily think of
reasonable extensions of the Standard Model where there are
significant new, possibly CP violating contributions.
(For specific examples, see \cite{DEKa,Ciuc,LoSo,BaSt}.)
Assuming that these new contributions do not induce CP violation
in decay, namely that $|\bar A_{\phi K_S}/A_{\phi K_S}|=1$ is
maintained, the new effects can be parameterized by
\be
{(\bar A_{\phi K_S}/A_{\phi K_S})^{\rm full}\over
(\bar A_{\phi K_S}/A_{\phi K_S})^{\rm SM}}=e^{2i\theta_A}.
\label{deftA}\ee
The result of such New Physics is that the asymmetries are
now modified as follows:
\be
\aPK=\sin2(\beta+\theta_d),\ \ \ \aFK=\sin2(\beta+\theta_d+\theta_A).
\label{NPaFK}\ee
Again, to test each of these predictions against the Standard Model
range (\ref{stwob}) requires modifications of order 50\%. The big
advantage of having $\Delta B=1$ effects is that (\ref{SMFP}) is
modified:
\be
\aPK\neq\aFK,
\label{NFFP}\ee
and that relatively small effects, of order 10\%, can lead to
an observable failure of (\ref{SMFP}). Therefore, measurements of
CP asymmetries in decays of $B^0$ that are suppressed by either
being FCNC processes or by small CKM angles, while experimentally
challenging, might provide exceptionally sensitive probes of
New Physics.
 
\section{$D$ Physics: CP Violation in $D-\bar D$ Mixing}
 
The best bound on $D-\bar D$ mixing comes from measurements of $D^0\ra
K^+\pi^-$.\cite{Dexp} However, these bounds are still orders of
magnitude above the Standard Model prediction for the mixing. If the
value of $\Delta m_D$ is anywhere close to present bounds, it should be
dominated by new physics. Then, new CP violating phases may play an
important role in $D-\bar D$ mixing. For example, new CP violating phases 
are expected in various supersymmetric models.\cite{NiSe,LNSb,CKLN}
 
The only type of CP violation that is likely to be relevant in
the experimental search for $D-\bar D$ mixing is in the interference
between decays with and without mixing:
 
(i) The decay $D^0\ra K^+\pi^-$ proceeds via the quark subprocess
$c\ra d\bar su$. Within the SM, this process is dominated by
doubly Cabibbo suppressed (DCS) tree amplitudes. It is very difficult,
if not impossible, for diagrams involving new physics to
contribute to this decay
comparably to the $W$-mediated diagram. Consequently, $D^0\ra K^+\pi^-$
is dominated by a single weak phase, $\arg(V_{cd}V_{us}^*)$, and it is
safe to neglect CP violation in decay.
 
(ii) If $\Delta m_D$ is close to present bounds, then it is clearly
dominated by new physics, $M_{12}\gg M_{12}^{\rm SM}$. On the other
hand, there is no reasonable type of new physics that could enhance
$\Gamma_{12}$ by orders of magnitude, so that very likely
$\Gamma_{12}\sim\Gamma_{12}^{\rm SM}$. Therefore, if $\Delta m_D$
is close to present bounds, it is safe to assume that $\Im(\Gamma_{12}/
M_{12})\ll1$ and neglect CP violation in mixing.
 
(iii) Within the Standard Model, both the mixing amplitude for neutral
$D$ mesons and the decay amplitude for $D\ra K\pi$ occur through
processes that involve, to a very good approximation, quarks of
the first two generation only. Therefore, the relative weak phase
between the mixing and decay amplitudes is extremely small. However,
most if not all extensions of the Standard Model that allow
$\Delta m_D$ close to the limit involve new CP violating phases.
In these models, the relative phase between the mixing amplitude
and the decay amplitude is usually unconstrained and would
naturally be expected to be of $\O(1)$. CP violation in the
interference between decays with and without mixing could then be
a large effect.
 
To understand the consequences of this situation, we introduce
the two quantities
\bea
\lambda_{K^+\pi^-}&=&\left({q\over p}\right)_D
{\bar A_{K^+\pi^-}\over A_{K^+\pi^-}},\nonumber\\
\lambda_{K^-\pi^+}&=&\left({q\over p}\right)_D
{\bar A_{K^-\pi^+}\over A_{K^-\pi^+}}.
\label{deflambdaD}\eea
Our discussion above of CP violation has the following implications:
Since CP violation in decay is negligible,
$|A_{K^+\pi^-}/\bar A_{K^-\pi^+}|=|\bar A_{K^+\pi^-}/A_{K^-\pi^+}|=1$.
Since CP violation in mixing is negligible, $|(q/p)_D|=1$. Then
\be
|\lambda_{K^+\pi^-}^{-1}|=|\lambda_{K^-\pi^+}|\equiv|\lambda|.
\label{equallam}\ee
Furthermore, since it experimentally known that
$\Delta m_D\ll\Gamma_D$ and $\Delta\Gamma_D\ll\Gamma_D$, we have
$|\lambda_{K^-\pi^+}|\ll1$. If $\Delta m_D$ is close to the bound then
$\Delta\Gamma_D\ll\Delta m_D$.
 
The result of this discussion is the following form for the
(time dependent) ratio between the DCS and Cabibbo-allowed decay rates:
\bea
{\Gamma[D^0(t)\ra K^+\pi^-)]\over\Gamma[D^0(t)\ra K^-\pi^+)]}&=&
|\lambda|^2+{(\Delta m_D)^2\over4}t^2+\Im(\lambda_{K^+\pi^-}^{-1})t,
\nonumber\\
{\Gamma[\bar D^0(t)\ra K^-\pi^+)]\over\Gamma[\bar D^0(t)\ra K^+\pi^-)]}
&=& |\lambda|^2+{(\Delta m_D)^2\over4}t^2+\Im(\lambda_{K^-\pi^+})t.
\label{DCStoCA}\eea
This form is valid for time $t$ not much larger than ${1\over\Gamma_D}$.
As concerns the linear term, there are four possible situations:
\begin{description}
\item[$1.$] {$\Im(\lambda_{K^+\pi^-}^{-1})=\Im(\lambda_{K^-\pi^+})=0$:
both strong and weak phases play no role in these processes.}
\item[$2.$] {$\Im(\lambda_{K^+\pi^-}^{-1})=\Im(\lambda_{K^-\pi^+})\neq0$:
weak phases play no role in these processes. There is a different strong
phase shift in $D^0\ra K^+\pi^-$ and $D^0\ra K^-\pi^+$.
(The strong phase shifts were calculated within two hadronic models
and found to be small.\cite{BrPa})}
\item[$3.$] {$\Im(\lambda_{K^+\pi^-}^{-1})=-\Im(\lambda_{K^-\pi^+})
\neq0$:
strong phases play no role in these processes. CP violating phases
affect the mixing amplitude.}
\item[$4.$] {$|\Im(\lambda_{K^+\pi^-}^{-1})|\neq
|\Im(\lambda_{K^-\pi^+})|$:
both strong and weak phases play a role in these processes.}
\end{description}
 
The linear term could be a problem for experiments: if the phase is such
that the interference is destructive, it could partially cancel the
quadratic term in the relevant range of time, thus weakening the
experimental sensitivity to mixing.\cite{BNS}
On the other hand, if the mixing amplitude is smaller than the DCS one,
the interference term may signal mixing even if the pure mixing
contribution is below the experimental sensitivity.\cite{WolfD}
 
\section{$K$ Physics: $K_L\ra\pi^0\nu\bar\nu$}
 
$K_L\ra\pi^0\nu\bar\nu$ is a very useful probe of CP violation:\cite{Litt}
 
(i) {\it It is dominated by short distance contributions.}
There is a hard GIM suppression of $\O(\Lambda_{\rm QCD}^2/m_c^2)$
between the long distance contribution and the charm mediated short
distance one (which by itself is small). It makes long distance
contributions negligibly small. QCD corrections are known to 
NLO\cite{BuBu,Bura} and electroweak corrections were calculated
to two loops in the large $m_t$ limit.\cite{BuBuEW}
 
(ii) {\it $\langle\pi|(\bar sd)_{V-A}|K\rangle$ is known.}
This matrix element
is a current operator that is much simpler than the four quark operators
that are relevant to other rare processes such as $\Delta m_K$ and
$\epsK$. Moreover, it is related by isopsin symmetry to
$\langle\pi|(\bar su)_{V_A}|K\rangle$ which is {\it measured} in
$K^+\ra\pi^0e^+\nu$ decay. The isospin breaking corrections
were calculated.\cite{MaPa}
 
(iii) {\it The decay is purely CP violating.}
In general, three body final states are not CP eigenstates.
However, in this case, if neutrinos are purely left-handed,
the final state is almost purely CP-even, with only $\O(m_K^2/m_Z^2)$
CP-odd component. Thus, the decay violates CP. (An exception
to this statement arises in models with lepton-number violating $K$
decays, where final states $\pi^0\nu_i\bar\nu_j$, $i\neq j$, could
dominate.\cite{GrNi})
 
(iv) {\it The required CP violation is dominated by interference
between decays with and without mixing.} It is experimentally
known that $|q/p|=1+\O(10^{-3})$. It is theoretically estimated 
that $|\bar A/A|=1+\O(10^{-5})$. In contrast, the effects
of CP violation in the interference of decays with and without
mixing ($\Im\lambda\neq0$) are expected to be of $\O(1)$.
 
\begin{figure}
\centerline{$(a)$}
\centerline{
\psfig{file=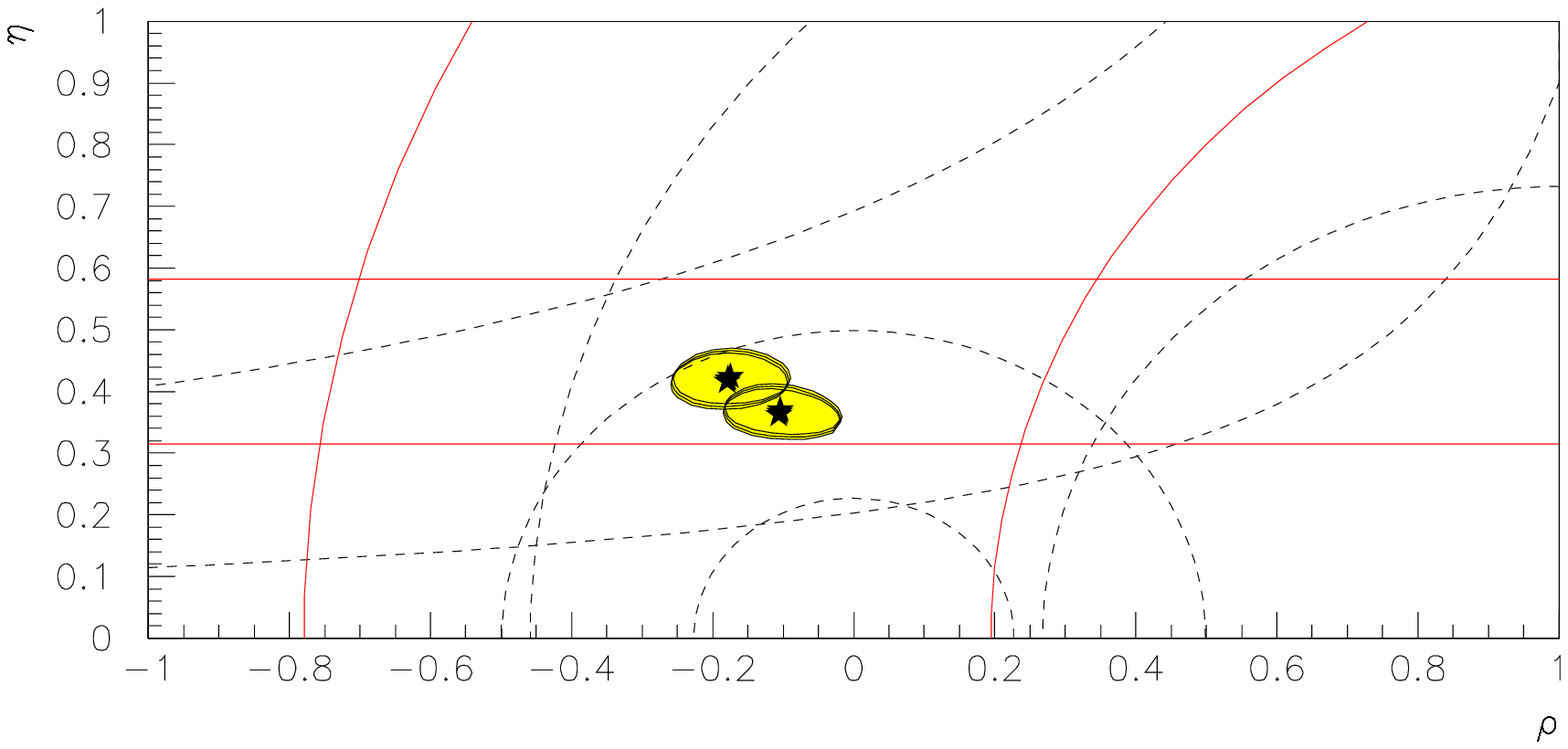,width=370pt,bbllx=0pt,bblly=410pt,bburx=612pt,bbury=660pt
}}
\centerline{$(b)$}
\centerline{
\psfig{file=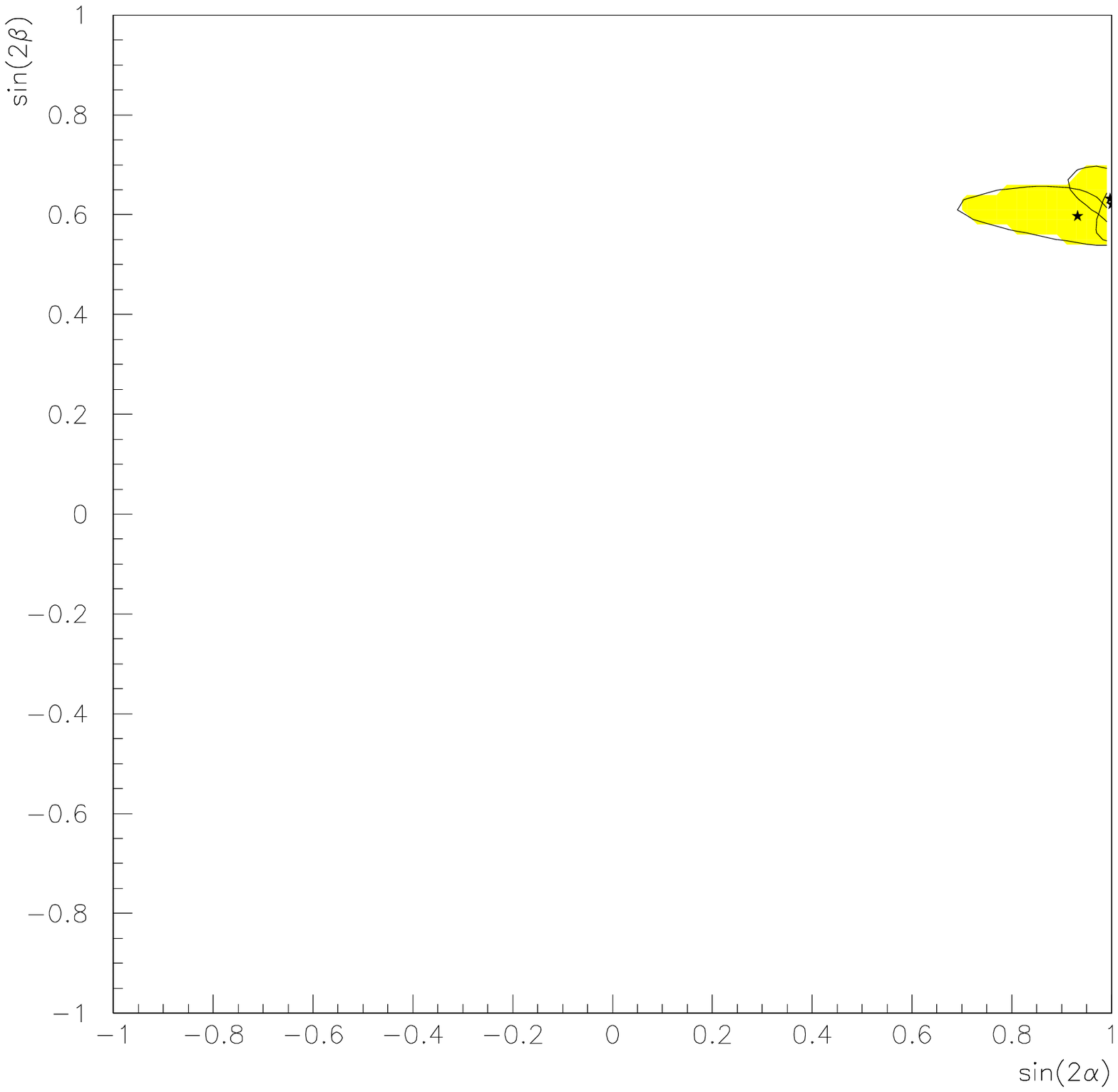,width=320pt,bbllx=0pt,bblly=160pt,bburx=612pt,bbury=653pt
}}
\caption{The effect of the $K\ra\pi\nu\bar\nu$ measurements
on the constraints $(a)$ in the $\rho-\eta$ plane and
$(b)$ in the $\sin2\alpha-\sin2\beta$ plane.
We use BR($K^+\ra\pi^+\nu\bar\nu)=(1.0\pm0.1)\times10^{-10}$,
BR($K_L\ra\pi^0\nu\bar\nu)=(3.0\pm0.3)\times10^{-11}$,
and $|V_{cb}|=0.039\pm0.02$. (These are the hypothetical ranges
used in ref. (65).) For all other constraints we
use present data.
\label{kpnnRE}}
\end{figure}

As a result of these special features, the $K_L\ra\pi^0\nu\bar\nu$
decay is theoretically clean to the level of $10^{-3}$.
The theoretical cleanliness (features (i) and (ii) above)
is also valid for the $K^+\ra\pi^+\nu\bar\nu$ decay.
This mode is, however, not CP violating. (Recently, the
first experimental evidence for this decay has been announced
by the E787 collaboration.\cite{Stew}) The combination of the
two decay modes provide a very clean determination of
the angle $\beta$ of the unitarity triangle.\cite{BuBub}
The cleanliness is comparable to that of the determination of
$\beta$ from the CP asymmetry in $B\ra\psi K_S$.
The constraints on the CKM parameters are demonstrated in
fig. \ref{kpnnRE}.
 
Model independently, we get a clean determination of $\theta_K$,
the relative phase between the $K-\bar K$ mixing amplitude and
the $s\ra d\nu\bar\nu$ decay amplitude:\cite{GrNi}
\be
a_{\pi\nu\bar\nu}\equiv{\Gamma(K_L\ra\pi^0\nu\bar\nu)\over
\Gamma(K^+\ra\pi^+\nu\bar\nu)}=\sin^2\theta_K.
\label{cleanK}\ee
Eq. (\ref{cleanK}) together with the experimental upper 
bound,\cite{Adle,Stew} 
$BR(K^+\ra\pi^+\nu\bar\nu)\leq2.4\times10^{-9}$,
give a model independent bound,\cite{GrNi}
\be
BR(K_L\ra\pi^0\nu\bar\nu)\leq1.1\times10^{-8},
\label{MIkpnn}\ee
which is more than two orders of magnitude stronger than
the new direct experimental bound from KTeV,\cite{StewH}
\be
BR(K_L\ra\pi^0\nu\bar\nu)\leq1.8\times10^{-6}.
\label{KTeV}\ee
 
The $K\ra\pi\nu\bar\nu$ decays are useful in probing CP violation
beyond the Standard Model.\cite{BGT,CDM,GrNi,Burd,BKS}
The bound (\ref{MIkpnn}) is still about three orders of
magnitude above the Standard Model prediction,\cite{BuraK}
$BR(K_L\ra\pi^0\nu\bar\nu)=(2.8\pm1.7)\times10^{-11}$,
leaving plenty of room for new physics. The $\epsK$ constraints
on CP violation in $K-\bar K$ mixing imply that such new physics
can only appear in the decay amplitude. For example, significant
new contributions to $s\ra d\nu\bar\nu$ with new CP violating
phases are possible in extensions of the quark sector.\cite{GrNi}
 
Finally, we would like to clarify one further point. In certain
superweak models, CP violation appears in processes that change
flavor by two units only, i.e. in mixing but not in decay amplitudes.
This leads to the prediction that the CP asymmetries in $K$
decays should be `universal', namely independent of the final state.
In particular, the CP asymmetry in $K\ra\pi\pi$ has been measured
(that is the $\epsK$ parameter) and is $\O(10^{-3})$. We learn
that if the ratio (\ref{cleanK}) is measured and found to be $\gg10^{-3}$
(or, equivalently at present, if
$BR(K_L\ra\pi^0\nu\bar\nu)\sim10^{-11}$, as predicted by the SM)
then superweak CP violation will be excluded. This situation is
sometimes described in the literature by the statement that
$K_L\ra\pi^0\nu\bar\nu$ will provide an unambiguous evidence for
direct CP violation. A similar conclusion will follow if
the asymmetries in, say, $B\ra\psi K_S$ and $B\ra\pi\pi$ are
found to be unequal.

\section{Supersymmetry}
 
\subsection{The Supersymmetric CP Problems}
 
A generic supersymmetric extension of the Standard Model contains a host
of new flavor and CP violating parameters. The requirement of consistency
with experimental data provides strong constraints on many of these
parameters. For this reason, the physics of flavor and CP violation
has had a profound impact on supersymmetric model building. A discussion
of CP violation in this context can hardly avoid addressing the flavor
problem itself.  Indeed, many of the supersymmetric models that we
analyze below were originally aimed at solving flavor problems.
 
As concerns CP violation, one can distinguish two classes of experimental
constraints. First, bounds on nuclear and atomic electric dipole moments
determine what is usually called the {\it supersymmetric CP problem}.
It involves effects that are flavor preserving and consequently appears
already in the minimal supersymmetric standard model (MSSM) with
universal sfermion masses and with the trilinear SUSY-breaking scalar
couplings proportional to the corresponding Yukawa couplings. In such a
constrained framework, there are two new physical phases beyond the
two phases of the Standard Model ($\delta_{\rm KM}$ and
$\theta_{\rm QCD}$),\cite{DGH,DiTh} usually denoted by $\phi_A$ and
$\phi_B$. In the more general case of non-universal soft terms there is
one independent phase $\phi_{A_{i}}$ for each quark and lepton flavor.
Moreover, complex off-diagonal entries in the sfermion
mass matrices may represent additional sources of CP violation.
 
The most significant effect of $\phi_A$ and $\phi_B$ is their
contribution to electric dipole moments (EDMs). In particular, the
present experimental bound, $d_N<1.1\times 10^{-25}e\, {\rm cm}$
\cite{smith,altarev} implies \cite{FPT}
\be
\left({100\, GeV\over \tilde m}\right )^2
\sin \phi_{A,B}\lsim10^{-2}{d_N\over10^{-25}\ e\, {\rm cm}},
\label{dipole}\ee
where $\tilde m$ represents the overall SUSY scale. Whether
the phases are small or squarks are heavy, a fine-tuning of order
$10^{-2}$ seems to be required, in general, to avoid too large a $d_N$.
This is the Supersymmetric CP Problem.
 
A second class of experimental constraints, involving
the physics of neutral mesons and, most importantly, the small
experimental value of $\epsK$, pose the {\it supersymmetric $\epsK$
problem}. The contribution to the CP violating $\epsK$ parameter in the
neutral $K$ system is dominated by diagrams involving $Q$ and $\bar d$
squarks in the same loop. A typical bound on the supersymmetric
parameters reads:\cite{GGMS}
\be
\left ({300 \ GeV\over\tilde m}\right)^2\left|
{(\delta m_Q^2)_{12} \over m_Q^2}{(\delta m_D^2)_{12}
\over m_D^2}\right|\sin\phi\lsim 0.5\times 10^{-7},
\label{epsKcon}\ee
where $\phi=\arg((\delta m_Q^2)_{12} (\delta m_D^2)_{12} )$,
and $(\delta m_{Q,D}^2)_{12}$ are the off diagonal entries in the
squark mass matrices in a basis where the down quark mass matrix
and the gluino couplings are diagonal. For dimensionless parameters
assuming their natural values of $O(1)$, the constraint
(\ref{epsKcon}) is generically violated by about seven orders of
magnitude. This is the supersymmetric $\epsK$ problem.
 
\subsection{Classes of Supersymmetric Models}
 
The supersymmetric flavor and CP problems have provided a very
significant input to supersymmetry model builders. Two scales
play an important role in supersymmetry: $\Lambda_S$, where
the soft supersymmetry breaking terms are generated, and $\Lambda_F$,
where flavor dynamics takes place.
 
Both supersymmetric CP problems are solved if, at the scale $\Lambda_S$,
the soft supersymmetry breaking terms are universal and the genuine SUSY
CP phases $\phi_{A,B}$ vanish. Then the Yukawa matrices represent
the only source of flavor and CP violation which is relevant in low
energy physics. This situation can naturally arise if
$\Lambda_S\ll\Lambda_F$, as in models where supersymmetry
breaking is mediated by the Standard Model gauge 
interactions.\cite{DNNS} In the simplest scenarios, 
the $A$-terms and the gaugino masses are generated by the 
same SUSY and $U(1)_R$ breaking source, leading to $\phi_A=0$. 
In specific models also $\phi_B=0$ in a similar way.\cite{DNeS,DNiS}
 
The most important implication of this type of boundary conditions
for soft terms, which we refer to as 
{\it exact  universality},\cite{DiGe,Saka} is the existence of the 
SUSY analogue of the GIM mechanism which operates in the SM. 
The CP violating phase of the CKM matrix can feed into the soft terms 
via Renormalization Group (RG) evolution only with a strong suppression 
from light quark masses.\cite{DGH,RoStru} The resulting phenomenology 
of CP violation is hardly distinguishable from the Standard Model.
 
When $\Lambda_F\lsim\Lambda_S$, we do not expect, in general, that flavor
and CP violation are limited to the Yukawa matrices. One way to suppress
CP violation would be to assume that CP is an approximate symmetry of the
full theory. In such a case, we expect also the SM phase
$\delta_{\rm KM}$ to be $\ll 1$. Then the standard box diagrams
cannot account for $\epsK$ which should arise from another
source. In supersymmetry with non-universal soft terms, the source could
be diagrams involving virtual superpartners, mainly squark-gluino box
diagrams. Let us call $(M_{12}^K)^{\rm SUSY}$
the supersymmetric contribution to the $K-\bar K$ mixing amplitude.
Then the requirements $\Re (M_{12}^K)^{\rm SUSY}\lsim\Delta m_K$
and $\Im(M_{12}^K)^{\rm SUSY}\sim\epsK\Delta m_K$ imply that the
generic CP phases are $\geq\O(\epsK)\sim 10^{-3}$. Then, somewhat
similar to the superweak scenario, all CP violating observables
(when defined appropriately) are characterized by a similar small
parameter. This situation implies many dramatic consequences,
{\it e.g.} $d_N$ just below or barely compatible with the present
experimental bound and, most striking, that
CP asymmetries in $B$ meson decays are small, perhaps
$\O(\epsK)$, rather than ${\cal O}(1)$ as expected in the SM.
 
Another option is to assume that, similarly to the Standard Model,
CP violating phases are large, but their effects are screened, possibly
by the same physics that explains the various flavor puzzles. This
usually requires Abelian or non-Abelian horizontal symmetries.
Two ingredients play a major role here: selection rules that come from
the symmetry and holomorphy of Yukawa and $A$-terms that comes from the
supersymmetry. With Abelian symmetries, the screening mechanism is
provided by {\it alignment},\cite{NiSe,LNSb} whereby the mixing matrices
for gaugino couplings have very small mixing angles, particularly for the
first two down squark generations. With non-Abelian symmetries, the
screening mechanism is {\it approximate universality}, where quarks of
the two light families fit into an irreducible doublet and are,
therefore, approximately degenerate.
\cite{DKL}$^-$\cite{HaMu}\ \cite{PoTo,BDH,CHM,Zurab,Raby}.
An extension of these ideas, aimed at screening the CP phases in the
$A$-terms, assumes that CP is a symmetry of the Lagrangian,\cite{RaNi}
spontaneously broken by the same fields that break the horizontal
symmetry. In general, it can be shown that non-universality of $A$-terms
and the requirement of $\O(1)$ CKM phase imply $\phi_A\gsim
\sin^6\theta_C$, leading to $d_N\gsim 10^{-28}e\ {\rm cm}$.
The minimal result can be reached only with almost triangular Yukawa
matrices, which can be achieved with Abelian flavor symmetries.
In models of non-Abelian symmetries, where the two light families are in
irreducible doublets, one does not expect such a structure and typically
the effective CP phases for light quarks are expected to be
$\gsim\sin^4\theta_C$.
 
As far as the third generation is concerned, the signatures of
Abelian and non-Abelian models are similar. In particular, they allow
observable deviations from the SM predictions for CP asymmetries in
$B$ decays. In some cases, non-Abelian models give relations between
CKM parameters and consequently predict strong constraints
on these CP asymmetries.
For the two light generations, only alignment allows
interesting effects. In particular, it predicts large CP violating
effects in $D-\bar D$ mixing.\cite{NiSe,LNSb}
 
Finally, it is possible that CP violating effects are suppressed because
squarks are heavy.\cite{DKS} 
If the masses of the first and second generations
squarks $m_i$ are larger than the other soft masses, $m_i^2\sim 100\,
\tilde m^2$ then the Supersymmetric CP problem is solved and the $\epsK$
problem is relaxed (but not eliminated).\cite{PoTo,DKL} This does not
necessarily lead to naturalness problems, since these two generations are
almost decoupled from the Higgs sector.
 
Notice though that, with the possible exception of $m_{\tilde b_R}^2$,
third family squark masses cannot naturally be much above $m_Z^2$.
If the relevant phases are of $O(1)$, the main contribution to $d_N$
comes from the third family via the two-loop induced three-gluon 
operator,\cite{Weintg} and it is roughly at the present experimental
bound  when $m_{\tilde t_{L,R}}\sim 100\ GeV$.
 
Models with the first two squark generations heavy have their own
signatures of CP violation in neutral meson mixing.\cite{CKLN}
The mixing angles relevant to $D-\bar D$ mixing are similar, in general,
to those of models of alignment (if alignment is
invoked to explain $\Delta m_K$ with $m^2_{Q,D}\lsim20\ TeV$).
However, as $\tilde u$ and $\tilde c$ squarks are heavy, the
contribution to $D-\bar D$ mixing is only about one to two orders
of magnitude below the experimental bound. This may lead to the
interesting situation that $D-\bar D$ mixing will first be observed
through its CP violating part.\cite{WolfD}
In the neutral $B$ system, $\O(1)$ shifts from the Standard Model
predictions of CP asymmetries in the decays to final CP eigenstates
are possible. This can occur even when the squarks masses of the third
family are $\sim1\ TeV$,\cite{CKNS} since now mixing angles can
naturally be larger than in the case of horizontal symmetries
(alignment or approximate universality).
 
To summarize, measurements of CP violation will provide us with an
excellent probe of the flavor and CP structure of supersymmetry.
This is clearly demonstrated in  Table (\ref{SUSYCP}).
 
\begin{table}[t]
\caption{CP violating observables in various classes
of Supersymmetric flavor models. $\theta_d$, $\theta_A$,
$\lambda_{K^-\pi^+}$ and $\theta_K$ are defined in eqs.
(21), (23), (26) and (29), respectively.\label{SUSYCP}}
\vspace{0.4cm}
\begin{center}
\begin{tabular}{|c|c|c|c|c|c|}
\hline & & & & \\
Model & ${d_N\over10^{-25}\ e\ {\rm cm}}$ & $\theta_d$ & $\theta_A$ &
${\Im(\lambda_{K^-\pi^+})\over|\lambda_{K^-\pi^+}|}$ &
$\theta_K$ \\ & & & & \\ \hline
Standard Model & $\lsim10^{-6}$ & $0$ & $0$ & $0$ & $\O(1)$ \\
Exact Universality & $\lsim10^{-6}$ & $0$ & $0$ & $0$ & =SM \\
Approximate CP & $\sim10^{-1}$ & $-\beta$ & $0$ & $\O(10^{-3})$ 
& $\O(10^{-3})$ \\
Alignment & $\gsim10^{-3}$ & $\O(0.2)$ & $\O(1)$ & 
$\O(1)$ & $\approx$SM \\
Approx. Universality & $\gsim10^{-2}$ & $\O(0.2)$ & $\O(1)$ & 
$0$ & $\approx$SM \\
Heavy Squarks & $\sim10^{-1}$ & $\O(1)$ & $\O(1)$ & 
$\O(10^{-2})$ & $\approx$SM \\
\hline
\end{tabular}
\end{center}
\end{table}
 
\section{Final Comments}
 
The unique features of CP violation are well demonstrated by
examining the CP asymmetry in $B\to\psi K_S$, $\aPK$,
and CP violation in $K_L\to\pi^0\nu\bar\nu$, $a_{\pi\nu\bar\nu}$.
Model independently, $\aPK$ measures the relative phase between
the $B-\bar B$ mixing amplitude and the $b\to c\bar cd$
decay amplitude (more precisely, the $b\to c\bar cs$ decay amplitude
times the $K-\bar K$ mixing amplitude),
while $a_{\pi\nu\bar\nu}$ measures the relative
phase between the $K-\bar K$ mixing amplitude and the
$s\to d\nu\bar\nu$ decay amplitude. We would like to emphasize
the following three points:
 
(i) {\it The two measurements are theoretically clean to better
than $\O(10^{-2})$.} Thus they can provide the most accurate
determination of CKM parameters. In particular, the theoretical
accuracy will be better than in the determination of $\sin\theta_C$
from $K\to\pi\ell\nu$.
 
(ii) {\it As concerns CP violation,
the Standard Model is a uniquely predictive model.}
In particular, it predicts that the seemingly unrelated
$\aPK$ and $a_{\pi\nu\bar\nu}$ measure the same parameter,
that is the angle $\beta$ of the unitarity triangle.
 
(iii) {\it In the presence of New Physics, there is in general
no reason for a relation between $\aPK$ and $a_{\pi\nu\bar\nu}$.}
Therefore, a measurement of both will provide a sensitive
probe of New Physics.
 
\section*{Acknowledgments}
I am grateful to Francesca Borzumati, Michael Dine,
Yuval Grossman and Nathan Seiberg for their help in preparing this talk.
Special thanks go to Stephane Plaszczynski and Marie-Helene Schune
for performing the CKM fit and preparing the plots for
my presentation.
YN is supported in part by the United States -- Israel Binational
Science Foundation (BSF), by the Israel Science Foundation,
and by the Minerva Foundation (Munich).

\end{document}